\documentclass[12pt]{article}
\usepackage{axodraw}
\usepackage{mathtext}        

\usepackage{amsmath,amssymb} 
\usepackage{epsfig}

\def\f21{{_2F_1}}

\def\al{\alpha}

\def\bp{\beta'}
\def\bt{\beta}

\def\ga{\gamma}

\newcommand{\be}{\begin{equation}}
\newcommand{\ee}{\end{equation}}
\newcommand{\bea}{\begin{eqnarray}}
\newcommand{\eea}{\end{eqnarray}}
\newcommand{\ba}{\begin{eqnarray*}}
\newcommand{\ea}{\end{eqnarray*}}

\newcommand{\rd}{{\rm d}}

\newcommand{\Fh}[2]{\,{}_#1F_#2}
\newcommand{\Fs}[3]{\!\!\left[\begin{array}{c}#1\,;\\#2\,;\end{array}#3\right]}

\newcommand{\Fpv}[2]{\Fs{#1}{#2}{\frac{p^2}{4m^2}}}

\newcommand{\FpR}[2]{\Fs{#1}{#2}{\frac{s_{12}-s_{13}}{m^2}}}

\newcommand{\Fep}[2]{\Fs{#1}{#2}{\frac{x}{2}}}

\newcommand{\Fxy}[2]{\Fs{#1}{#2}{\frac{\omega^2}{(\omega-z)(\omega-1)}}}

\newcommand{\Fca}[2]{\Fs{#1}{#2}{\frac{x^2}{4(x-1) }}}

\newcommand{\Fom}[2]{\Fs{#1}{#2}{1-\frac{m_1^2}{m_2^2}}}

\newcommand{\Fmm}[2]{\Fs{#1}{#2}{\frac{s_{13}}{4 \sigma}}}

\newcommand{\Fxk}[2]{\Fs{#1}{#2}{x^2}}
\newcommand{\Fpo}[2]{\Fs{#1}{#2}{\frac{-m^2}{\sigma} }}
\newcommand{\Fbb}[2]{\Fs{#1}{#2}{\frac{(x-3)(x+1)^3}{(x+3)(x-1)^3 }}}
\newcommand{\FXX}[2]{\Fs{#1}{#2}{\frac{x^2(x^2-9)^2}{(x^2+3)^3}}}
\newcommand{\Fwz}[2]{\Fs{#1}{#2}{\frac{w-z}{1-z}}}

\catcode`@=11
\newcount\@tempcntc
\def\@citex[#1]#2{\if@filesw\immediate\write\@auxout{\string\citation{#2}}\fi
  \@tempcnta\z@\@tempcntb\m@ne\def\@citea{}\@cite{\@for\@citeb:=#2\do
    {\@ifundefined
       {b@\@citeb}{\@citeo\@tempcntb\m@ne\@citea\def\@citea{,}{\bf ?}\@warning
       {Citation `\@citeb' on page \thepage \space undefined}}%
    {\setbox\z@\hbox{\global\@tempcntc0\csname b@\@citeb\endcsname\relax}%
     \ifnum\@tempcntc=\z@ \@citeo\@tempcntb\m@ne
       \@citea\def\@citea{,}\hbox{\csname b@\@citeb\endcsname}%
     \else
      \advance\@tempcntb\@ne
      \ifnum\@tempcntb=\@tempcntc
      \else\advance\@tempcntb\m@ne\@citeo
      \@tempcnta\@tempcntc\@tempcntb\@tempcntc\fi\fi}}\@citeo}{#1}}
\def\@citeo{\ifnum\@tempcnta>\@tempcntb\else\@citea\def\@citea{,}%
  \ifnum\@tempcnta=\@tempcntb\the\@tempcnta\else
   {\advance\@tempcnta\@ne\ifnum\@tempcnta=\@tempcntb \else \def\@citea{--}\fi
    \advance\@tempcnta\m@ne\the\@tempcnta\@citea\the\@tempcntb}\fi\fi}
\catcode`@=12
\begin{document}

%%%%%%%%%%%%%%%%%%%%%%%%%%%%%%%%%%%%%%%%%%%%%%%
%\begin{titlepage}
%=========== title page ======================
%\thispagestyle{empty}
%\onecolumn

\title{
\vskip-3cm{\baselineskip14pt
\centerline{\normalsize DESY~11--145\hfill ISSN~0418--9833}
\centerline{\normalsize NSF--KITP--11--127\hfill}
\centerline{\normalsize August 2011\hfill}}
\vskip1.5cm
Finding new relationships between hypergeometric functions by evaluating
Feynman integrals}

\author{Bernd A. Kniehl$^1$\thanks{On leave of absence from
II. Institut f\"ur Theoretische Physik, Universit\" at Hamburg,
Luruper Chaussee 149, 22761 Hamburg, Germany.},\
Oleg V. Tarasov$^2$\thanks{On leave of absence from
Joint Institute for Nuclear Research, 141980 Dubna (Moscow Region), Russia.} \\
\\
$^1$ Kavli Institute for Theoretical Physics, Kohn Hall,\\
University of Santa Barbara, CA~93106, USA \\
$^2$ II. Institut f\"ur Theoretische Physik, Universit\"at Hamburg,\\
Luruper Chaussee 149, 22761 Hamburg, Germany}

\date{}

\maketitle

\begin{abstract}
Several new relationships between hypergeometric functions are found by
comparing results for Feynman integrals calculated using different methods.
%By applying different methods for calculating the same Feynman 
%integrals we find new relationships between hypergeometric functions.
A new expression for the one-loop propagator-type integral with arbitrary
masses and arbitrary powers of propagators is derived in terms of only one
Appell hypergeometric function $F_1$.
From the comparison of this expression with a previously known one, a new
relation between the Appell functions $F_1$  and $F_4$ is found.
By comparing this new expression for the case of equal masses with another
known result, a new formula for reducing the $F_1$ function with particular
arguments to the hypergeometric function $_3F_2$ is derived.
By comparing results for a particular one-loop vertex integral
obtained using different methods, a new relationship between $F_1$ functions
corresponding to a quadratic transformation of the arguments is established.
Another reduction formula for the $F_1$ function is found by analysing the
imaginary part of the two-loop self-energy integral on the cut.
An explicit formula relating the $F_1$ function and the Gaussian hypergeometric
function $_2F_1$ whose argument is the ratio of polynomials of degree six is
presented.
\medskip


\end{abstract}

\newpage

\section{Introduction}

Radiative corrections to different physical quantities needed for the
comparison of theoretical predictions with experimental data to be collected
with the CERN Large Hadron Collider (LHC) and, in future, with an International
Linear Collider (ILC) and other colliders are expressed in terms of complicated
Feynman integrals.
In many cases, radiative corrections must be evaluated analytically to achieve
reliable accuracies in the calculations.
The difficulties in calculating Feynman integrals are usually related to the
fact that they depend on several kinematical scales, i.e.\ they are functions
of  several variables.

Nowadays, one of the most frequently used methods for calculating Feyman
integrals is based on the Mellin-Barnes integral representation
\cite{Boos:1990rg,Davydychev:1990jt,Davydychev:1990cq}.
In many cases, however, this method leads to complicated expressions in terms
of hypergeometric functions with many variables.
In order to calculate analytically integrals with several kinematical variables
and masses, new effective methods are to be developed.
Rather promising methods for analytic calculations of Feynman integrals are
based on recurrence relations.
These can be recurrence relations with respect to the exponent of a propagator
of the integral \cite{Kazakov:1983pk} or the parameter of the space-time
dimension \cite{Tarasov:1996br,Tarasov:2000sf,Tarasov:2006nk}.
As was already observed in the one-loop case, the solutions of dimensional
recurrences are combinations of hypergeometric functions
\cite{Tarasov:1996br,Tarasov:2000sf,Fleischer:2003rm}.
This is also true at the two-loop level \cite{Tarasov:2006nk}.

As was realized many years ago in Ref.~\cite{Regge:1967}, Feynman integrals are
generalized hypergeometric functions.
This conjecture was confirmed through the evaluations of specific Feynman
integrals.
Some results for Feynman integrals expressed in terms of hypergeometric
functions may be found in
Refs.~\cite{Boos:1990rg,%
Davydychev:1990jt,%
Davydychev:1990cq,%
Bollini:1972bi,%
Kalinowski:1982,%
InayatHussain:1987,%
Broadhurst:1987ei,%
JPA,%
Davydychev:1992mt,%
Broadhurst:1993mw,% 
Davydychev:1993ut,%
Berends:1993ee,%
Fleischer:1998dw,%
CabralRosetti:1998sp,%
Anastasiou:1999ui,%
Anastasiou:1999cx,% 
DavydychevKalmykov_Hgf,%
Davydychev:2005nf}.
%To obtain these results rather  different methods of calculation were developed. 
These results were obtained using rather different methods, e.g.\ 
by directly evaluating the integrals from their Feynman parameter
representations, by applying Mellin-Barnes integral representations, 
by solving recurrence relations, by making use of the negative-dimension
approach \cite{Halliday:1987an}, or by using spectral representations.

As a method for finding relations between hypergeometric functions, the authors
of Ref.~\cite{Srivastava_Karlsson} advocated the evaluation
of integrals reducible to hypergeometric functions by several
different methods and the comparison of the results thus obtained.
In this respect, the evaluation of Feynman integrals may be considered as a
rich source for finding relations between hypergeometric functions. 
New transformation and reduction formulae for hypergeometric functions were
derived by calculating Feynman integrals already a long time ago
\cite{InayatHussain:1987}.
Several new reduction relations for the Appell hypergeometric functions $F_1$
and $F_4$ obtained by comparing different results for the same Feynman integral
were presented in  Ref.~\cite{Shpot:2007bz}.

The analytic evaluation of Feynman integrals offers us a unique possibility to
find relations between hypergeometric functions which can be useful in many
other applications, far away from high-energy physics.
On the other hand, the problems emerging when evaluating Feynman integrals may
become interesting for mathematicians, and their participation in the solution
of these problems may lead to essential progress in the evaluation of Feynman
integrals.

Our paper organized as follows.
In section~\ref{sec:two}, we present a new result for a one-loop
propagator-type integral with arbitrary exponents of propagators and arbitrary
masses.
In section~\ref{sec:three}, 
%by comparing the result given in section 2 with the result 
%derived in \cite{Boos:1990rg}
a new formula for the reduction of the Appell function $F_4$ to the function
$F_1$ is derived.
Setting the masses in the result derived in section 2 to be equal and comparing
the outcome with a known result, a new formula for the reduction of the Appell
function $F_1$ to the hypergeometric function $_3F_2$ is obtained.
In section~\ref{sec:four}, from the results for the one-loop vertex-type
integral, a quadratic transformation formula for the Appell function $F_1$ is
derived.
In section~\ref{sec:five}, from the comparison of results for the imaginary
part of a two-loop self-energy integral obtained using different methods, a
formula for the reduction of the $F_1$ function to the Gauss hypergeometric
function with a complicated argument is obtained.
In section~\ref{sec:six}, we present a short summary of our results. 
%%%%%%%%%%%%%%%%%%%%%%%%%%%%%%%%%%%%%%%%%%%%%%%%%%%%%%%%%%%%%%%%%%%%

\section{New analytic expression for the one-loop propagator-type integral}
\label{sec:two}

%%%%%%%%%%%%%%%%%%%%%%%%%%%%%%%%%%%%%%%%%%%%%%%%%%%%%%%%%%%%%%%%%%%%

\begin{figure}[h]
\begin{center}
\includegraphics[scale=1.0]{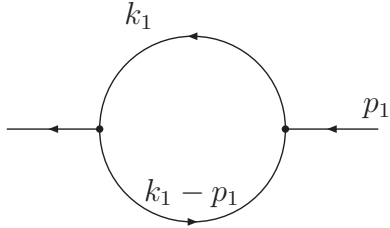}
\end{center}
\caption{Feynman diagram corresponding to the integral $I_{\nu_1 \nu_2}^{(d)}$.}
\end{figure}
In this section, we consider the evaluation of the one-loop propagator type
integral with arbitrary masses and arbitrary powers of propagators,
\begin{equation}
I_{\nu_1\nu_2}^{(d)}(m_1^2,m_2^2;~s_{12})=
\int \frac{d^dq}{i\pi^{d/2}}
\frac{1}{[(q-p_1)^2-m_1^2]^{\nu_1}[(q-p_2)^2-m_2^2]^{\nu_2}}.
\label{defineI2}
\end{equation}
Here and below, it is understood that the usual causal prescription of the
propagators is used, i.e.\ $1/[k^2-m^2] \leftrightarrow 1/[k^2-m^2+i0]$.
The Feynman diagram corresponding to this integral is presented in Figure~1.
By using the formula
\begin{equation}
\frac{1}{a^{\nu_1}b^{\nu_2}}=\frac{\Gamma(\nu_1+\nu_2)}
{\Gamma(\nu_1)\Gamma(\nu_2)} \int_0^1dx \frac{x^{\nu_1-1}(1-x)^{\nu_2-1}}
{[ax+b(1-x)]^{\nu_1+\nu_2} },
\end{equation}
the product of the propagator factors can be transformed to an integral over
Feynman parameters.
Changing the integration momentum as $q=t + p_2 +x(p_1-p_2)$ and applying the
formula 
\begin{equation}
\int \frac{d^dt}{i \pi^{d/2}} \frac{1}{(t^2-M^2)^{\nu}} =
(-1)^{\nu} \frac{\Gamma\left(\nu - \frac{d}{2}\right)}
 {(M^2)^{\nu-\frac{d}{2}}\Gamma(\nu)},
\end{equation}
we obtain the following representation for the integral of Eq.~(\ref{defineI2}):
\begin{equation}
I_{\nu_1\nu_2}^{(d)}(m_1^2,m_2^2;~s_{12})=
(-1)^{\nu_1+\nu_2}\frac{\Gamma\left(\nu_1+\nu_2-\frac{d}{2}\right)}
{\Gamma(\nu_1)\Gamma(\nu_2)}
\int_0^1\frac{dx~x^{\nu_1-1}(1-x)^{\nu_2-1}}
{[s_{12}x^2+x(m_1^2-m_2^2-s_{12})+m_2^2]^{\nu_1+\nu_2-\frac{d}{2}}}.
\end{equation}
Representing the quadratic polynomial in the denominator as
\begin{equation}
s_{12}x^2+x(m_1^2-m_2^2-s_{12})+m_2^2=m_2^2(1-x_1x)(1-x_2x),
\end{equation} 
and then comparing our integral with the integral representation
for the Appell function \cite{ApKdF},
\begin{equation}
F_1(a,b, b'; c;~w,z)=\frac{\Gamma(c)}{\Gamma(a)\Gamma(c-a)}
\int_0^1  \frac{du~u^{a-1}(1-u)^{c-a-1}}
{(1-uw)^b(1-uz)^{b'}},
\label{F1_int_repr}
\end{equation}
the following result follows:
\begin{eqnarray}
&&I_{\nu_1\nu_2}^{(d)}(m_1^2,m_2^2;~s_{12})=
\frac{(-1)^{\nu_1+\nu_2}\Gamma\left(\nu_1+\nu_2-\frac{d}{2}\right)}
{\Gamma(\nu_1+\nu_2) (m_2^2)^{\nu_1+\nu_2-d/2}}
\nonumber
\\
&&~~~~~~~~~~~~~~~~~~~
\times
F_1\left(\nu_1,\nu_1+\nu_2-\frac{d}{2},\nu_1+\nu_2-\frac{d}{2};
\nu_1+\nu_2;x_{-},x_{+}\right),
\label{I2intermsF1}
\end{eqnarray}
where
\begin{eqnarray}
&&x_{\pm} = \frac{1+x-y \pm \sqrt{\Lambda(1,x,y)}}{2},
\nonumber \\
&&x=\frac{s_{12}}{m_2^2},~~~~~~y=\frac{m_1^2}{m_2^2},
\label{x_pm}
\end{eqnarray}
with 
\begin{equation}
\Lambda(x,y,z) = (x-y-z)^2-4yz.
\label{eq:lambda}
\end{equation}
%\begin{eqnarray}
%&&x_1=\frac{s_{12}+m_2^2-m_1^2-\sqrt{\Delta}}{2m_2^2},~~~~~
%x_2=\frac{s_{12}+m_2^2-m_1^2+\sqrt{\Delta}}{2m_2^2},
%\nonumber 
%\\

For the particular case $\nu_1=\nu_2=1$, an expression for the propagator-type
integral in terms of the Appell function $F_1$ was given in
Refs.~\cite{Kalinowski:1982,Shpot:2007bz,Kaneko:2011ym}.

The integral $I_{\nu_1\nu_2}^{(d)}(m_1^2,m_2^2;~s_{12})$ is symmetric with respect to
the change $\nu_1,m_1^2 \leftrightarrow \nu_2, m_2^2$.
%The right hand side of Eq.(\ref{I2intermsF1}) is also symmetric 
To understand this symmetry, we first observe that, under the change
$m_1^2 \leftrightarrow m_2^2$, the  arguments of the Appell function $F_1$ in
Eq.~(\ref{I2intermsF1}) transform as
\begin{equation}
x_{-} \rightarrow \frac{x_{-}}{x_{-}-1},~~~~~~
x_{+} \rightarrow \frac{x_{+}}{x_{+}-1},
\end{equation}
and then, applying the formula (see, for example,
Refs.~\cite{ApKdF,Erdely,Prudnikov:1986})
\begin{eqnarray}
F_1\bigl( \al,\bt,\bp,\ga,w,z\bigl) 
&=& (1-w)^{-\bt}(1-z)^{-\bp}
    F_1\bigl(\ga-\al,\bt,\bp,\ga,\frac{w}{w-1},\frac{z}{z-1} \bigl), 
\end{eqnarray}
we return to the initial expression on the right-hand side of
Eq.~(\ref{I2intermsF1}).

\vspace{0.5cm}

%%%%%%%%%%%%%%%%%%%%%%%%%%%%%%%%%%%%%%%%%%%%%%%%%%%%%%%%%%%%%%%%%%%%%%%

\boldmath
\section{Relations between the $F_1$ function and other hypergeometric
functions}
\unboldmath
\label{sec:three}

%%%%%%%%%%%%%%%%%%%%%%%%%%%%%%%%%%%%%%%%%%%%%%%%%%%%%%%%%%%%%%%%%%%%%%%

In Ref.~\cite{Boos:1990rg}, by exploiting the Mellin-Barnes integral 
representation, the following analytic expression for the considered integral
was derived:
\begin{eqnarray}
&&I_{\nu_1\nu_2}^{(d)}(m_1^2,m_2^2;~s_{12})=
\frac{(-1)^{\nu_1+\nu_2}}
{(m_2^2)^{\nu_1+\nu_2-d/2}}
\left\{ \frac{\Gamma\left( \frac{d}{2}-\nu_1\right)
\Gamma \left( \nu_1+\nu_2-\frac{d}{2}\right)}
{\Gamma \left( \frac{d}{2} \right) \Gamma(\nu_2)}
\right.
\nonumber
\\
&&~~\times 
F_4\left( \nu_1,\nu_1+\nu_2-\frac{d}{2};
\frac{d}{2},\nu_1-\frac{d}{2}+1;
\frac{s_{12}}{m_2^2},\frac{m_1^2}{m_2^2} \right)
\nonumber
\\
&&\left.
+\left(\frac{m_1^2}{m_2^2}\right)^{\frac{d}{2}-\nu_1}
\frac{\Gamma\left(\nu_1-\frac{d}{2}\right)}
{\Gamma(\nu_1)}
F_4\left(\nu_2,\frac{d}{2};\frac{d}{2},\frac{d}{2}-\nu_1+1;
\frac{s_{12}}{m_2^2},\frac{m_1^2}{m_2^2}\right)
\right\}.
\label{I2intermsF4}
\end{eqnarray}
A hypergeometric representation in terms of Lauricella functions for the
one-loop integrals corresponding to diagrams with an arbitrary number of
external legs was presented in
Refs.~\cite{Davydychev:1990jt,Davydychev:1990cq}.

Comparing Eqs.~(\ref{I2intermsF1}) and (\ref{I2intermsF4}), we arrive at the
following relation:
\begin{eqnarray}
&&
F_1\left(\nu_1,\nu_1+\nu_2-\frac{d}{2},\nu_1+\nu_2-\frac{d}{2};
\nu_1+\nu_2;~x_{-},x_{+}\right)=
\nonumber
\\
&&~~ \frac{\Gamma\left( \frac{d}{2}-\nu_1\right)
\Gamma( \nu_1+\nu_2)}
{\Gamma \left( \frac{d}{2} \right) \Gamma(\nu_2)}
~
F_4\left( \nu_1,\nu_1+\nu_2-\frac{d}{2};
\frac{d}{2},\nu_1-\frac{d}{2}+1; ~x,y \right)
\nonumber
\\
&&
+y^{\frac{d}{2}-\nu_1}
\frac{\Gamma\left(\nu_1-\frac{d}{2}\right)
\Gamma(\nu_1+\nu_2)
}
{\Gamma(\nu_1) \Gamma\left(\nu_1+\nu_2-\frac{d}{2}\right)}
F_4\left(\nu_2,\frac{d}{2};\frac{d}{2},
\frac{d}{2}-\nu_1+1;~x,y\right).
\label{F1intermsF4}
\end{eqnarray}
Here, $x_{\pm}$ are given by Eq.~(\ref{x_pm}).
With the help of the relation given in Ref.~\cite{Bailey:1964} (see p.~102),   
\begin{equation}
F_4\left(\alpha,\beta; \gamma, \beta;
\frac{-x}{(1-x)(1-y)}, \frac{-y}{(1-x)(1-y)}  \right)= 
[(1-x)(1-y)]^{\alpha} 
F_1(\alpha,\gamma-\beta,1+\alpha-\gamma, \gamma;
x, xy),
\end{equation}
the second Appell functions $F_4$ on the right-hand side of
Eq.~(\ref{F1intermsF4}) may be expressed in terms of the Appell function $F_1$.
Therefore, the following relation holds:
\begin{eqnarray}
&&
F_4(\alpha,\beta,\beta',\alpha-\beta'+1; x,y)
\nonumber \\
&&~~~~~~~
=\frac{ \Gamma(\beta') \Gamma(\beta-\alpha+\beta') }
      { \Gamma(\beta'-\alpha) \Gamma(\beta+\beta') }
 F_1(\alpha,\beta,\beta,\beta+\beta';x_{-},x_{+})
\nonumber \\
&&~~~~~~~
 -\frac{\Gamma(\beta') \Gamma(\beta-\alpha+\beta') 
        \Gamma(\alpha-\beta')}
       {\Gamma(\alpha) \Gamma(\beta) \Gamma(\beta'-\alpha)
          }~ y^{\beta'-\alpha}~(x_{+}-x)^{\alpha-\beta-\beta'}
\nonumber \\
&&~~~~~~~\times
    F_1\left(\beta-\alpha+\beta',1-\alpha,\beta,
     \beta'-\alpha+1; 
    \frac{x-x_{-}}{x},
    \frac{x-x_{-}}{x-x_{+}}\right).
\end{eqnarray}

In Ref.~\cite{Davydychev:1990cq}, an expression for the integral
$I_{\nu_1\nu_2}^{(d)}(m_1^2,m_2^2;~s_{12})$ in terms of the Kamp\'e de F\'eriet
function was derived:
\begin{eqnarray}
&&
I_{\nu_1\nu_2}^{(d)}(m_1^2,m_2^2;~s_{12})
=(-1)^{\nu_1+\nu_2}(m_2^2)^{\frac{d}{2}-\nu_1-\nu_2}
\frac{\Gamma\left(\nu_1+\nu_2-\frac{d}{2}\right)}
{\Gamma(\nu_1+\nu_2)}
\nonumber \\
&&~~~~~~~~
~\times~F_{1;0;0}^{2;1;0}\left[ 
\begin{matrix} (\nu_1+\nu_2-d/2:1,1),(\nu_1:1,1):(\nu_2:1)\\
(\nu_1+\nu_2:2,1)
\end{matrix}
\left| \frac{s_{12}}{m_2^2},1-\frac{m_1^2}{m_2^2}\right.
\right].
\end{eqnarray}
Comparing this relation with Eq.~(\ref{I2intermsF1}), we obtain the following
reduction formula:
\begin{eqnarray}
F_{1;0;0}^{2;1;0}\left[ 
\begin{matrix} (\alpha:1,1),(\nu_1:1,1):(\nu_2:1)\\
(\nu_1+\nu_2:2,1)
\end{matrix}
\Biggl| x,y \Biggr.
\right]=
F_1(\nu_1,\alpha,\alpha,\nu_1+\nu_2; ~z_{-},z_{+}),
\end{eqnarray}
where
\begin{equation}
z_{\pm}=\frac{x+y\pm\sqrt{(x+y)^2-4x}}{2}.
\end{equation}

For the case of equal masses $m_2^2=m_1^2=m^2$, the following expression was
derived in Ref.\cite{Boos:1990rg}:
\begin{eqnarray}
&&I_{\nu_1\nu_2}^{(d)}(m^2,m^2;~s_{12})=(-1)^{\nu_1+\nu_2}
(m^2)^{d/2-\nu_1-\nu_2} 
\nonumber
\\
&&~~~~~~\times
\frac{\Gamma\left(\nu_1+\nu_2-\frac{d}{2}\right)}
{\Gamma(\nu_1+\nu_2)}
\Fh32\Fpv{\nu_1,\nu_2,\nu_1+\nu_2-\frac{d}{2} }
{\frac{\nu_1+\nu_2}{2}, \frac{\nu_1+\nu_2+1}{2}}.
\end{eqnarray}
Comparing this formula with Eq.~(\ref{I2intermsF1}) taken at $m_1^2=m_2^2=m^2$,
we obtain:
\begin{equation}
F_1\left(\alpha,\beta, \beta; \gamma;~x-\sqrt{x^2-2x},
~x+\sqrt{x^2-2x} ~ \right)=
\Fh32\Fep{\alpha,\gamma-\alpha,\beta}
{\frac{\gamma}{2},\frac{\gamma+1}{2}},
\end{equation} 
which may be rewritten as:
\begin{equation}
F_1\left(\alpha,\beta, \beta; \gamma;~x,
~\frac{x}{x-1} ~ \right)=
\Fh32\Fca{\alpha,\gamma-\alpha,\beta}
{\frac{\gamma}{2},\frac{\gamma+1}{2}}.
\label{new_f1_f32}
\end{equation} 
We verified numerically the correctness of this relation setting $\alpha=1/2$,
$\beta=2$, $\gamma=3/2$, and $x=-1/2$, and keeping 600 valid digits in the
calculations performed using the computer algebra system Maple.

To the best of our knowledge, there is no such a relation in the mathematical
literature, i.e.\ Eq.~(\ref{new_f1_f32}) extends the number of known reduction
formulas for the Appell function $F_1$.
In Ref.~\cite{Ismail}, a relation between $F_1$ with the same arguments and
the Gauss hypergeometric function $_2F_1$ is given.
That relation corresponds to a particular case of our Eq.~(\ref{new_f1_f32}),
taken at $\gamma=2\alpha$.
We would like to recall that the only known transformation of the Appell
function $F_1$ to the hypergeometric function $_3F_2$ was known for the case
when $x=-y$.

For the Appell hypergeometric function $F_1$, the following relation holds:
\begin{equation}
F_1(\alpha,\beta,\beta; \gamma;x,-x)=
\Fh32\Fxk{\frac{\alpha}{2}, \frac{\alpha+1}{2},\beta}
{\frac{\gamma}{2},\frac{\gamma+1}{2}}.
\end{equation}
In the case when $s_{12}=m_1^2-m_2^2$, the relation $x_{+}=-x_{-}$ holds.
Therefore, the integral $I_{\nu_1\nu_2}^{(d)}(m_1^2,m_2^2;~s_{12})$ may be expressed
in terms of the hypergeometric function $_3F_2$ as
\begin{eqnarray}
&&\left.I_{\nu_1\nu_2}^{(d)}(m_1^2,m_2^2;~s_{12})
\right|_{s_{12}=m_1^2-m_2^2}
\nonumber \\
&&=\frac{(-1)^{\nu_1+\nu_2}\Gamma\left(\nu_1+\nu_2-\frac{d}{2}\right)}
{\Gamma(\nu_1+\nu_2) (m_2^2)^{\nu_1+\nu_2-d/2}}
\Fh32\Fom{\frac{\nu_1}{2},\frac{\nu_1+1}{2},\nu_1+\nu_2-\frac{d}{2}}
{\frac{\nu_1+\nu_2}{2},\frac{\nu_1+\nu_2+1}{2}}.
\end{eqnarray}
This formula demonstrates that simplifications of Feynman integrals may also
take place for specific values of masses or momenta that are more general than
just zero or on-shell.

%%%%%%%%%%%%%%%%%%%%%%%%%%%%%%%%%%%%%%%%%%%%%%%%%%%%%%%
%\section{Evaluation of the one-loop vertex type integral
% $I_3^{(d)}(0,m^2,0;~0,s_{13},s_{12})$}
\boldmath
\section{Quadratic transformation for the Appell function $F_1$}
\unboldmath
\label{sec:four}

%%%%%%%%%%%%%%%%%%%%%%%%%%%%%%%%%%%%%%%%%%%%%%%%%%%%%%%
\begin{figure}[h]
\begin{center}
\includegraphics[scale=1.1]{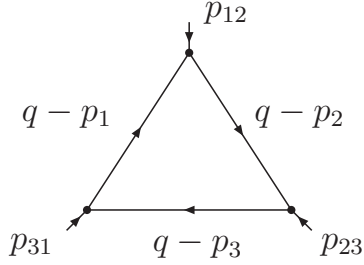}
\end{center}
\caption{Feynman diagram corresponding to the integral $I_3^{(d)}$.}
\end{figure}
In this section, we find relations for the Appell function $F_1$ by comparing
the results of different calculations of the one-loop vertex-type integral 
\begin{equation}
I_3^{(d)}(m_j^2,m_k^2,m_l^2;p_{kl},p_{jl},p_{jk})\!=\!
\int \frac{d^d q}{i \pi^{{d}/{2}}}
\frac{1}{[(q-p_j)^2\!-\!m_j^2]
         [(q-p_k)^2\!-\!m_k^2]
         [(q-p_l)^2\!-\!m_l^2]},
\end{equation}
corresponding to the Feynman diagram shown in Figure~2.
We consider this integral with a particular set of arguments, namely
$I_3^{(d)}(0,m^2,0;~0,s_{13},s_{12})$.
In the case when $s_{12}\leq m^2$ and $s_{13}\leq m^2$, from its
representation as an integral over Feynman parameters, we derive the following
analytic expression:
\begin{eqnarray}
&&I_3^{(d)}(0,m^2,0;~0,s_{13},s_{12})
%\nonumber
%\\
%&&
=\int_0^1\int_0^1 \frac{- dx_1dx_2~\Gamma\left(3-\frac{d}{2}\right)x_1}
{[ s_{13}x_1-s_{13} +x_2(m^2+s_{13}-s_{12}-x_1s_{13}
+x_1s_{12})]^{3-\frac{d}{2}}}
\nonumber
\\
&&
\nonumber\\
&&=\frac{1}{m^2}~I_{11}^{(d)}(0,m^2;~0)
~F_1\left(1,1,2-\frac{d}{2}, \frac{d}{2};
  \frac{s_{12}-s_{13}}{m^2}, \frac{s_{12}}{m^2}\right)
\nonumber
\\
&&~~~~~~~~~~~~~~~~~~~
-\frac{I_{11}^{(d)}(0,0;~s_{13}) }
{m^2}\Fh21\FpR{1,\frac{d-2}{2}}{d-2},
\label{AppellF1}
\end{eqnarray}
where  $F_1$ is the Appell hypergeometric function \cite{ApKdF} defined by the
series
\begin{equation}
F_1(a,b,b';c;w,z) = \sum_{k,l=0}^{\infty}
\frac{(a)_{k+l}(b)_k (b')_l}{(c)_{k+l}} 
\frac{w^kz^l}{k! l!},
\end{equation}
and $(a)_k=\Gamma(a+k)/\Gamma(a)$ is the so-called Pochhammer symbol.
In our case, the Appell function has a rather simple integral representation,
viz.\
\begin{equation}
F_1\left(1,1,2-\frac{d}{2},\frac{d}{2};x,y\right)=
\frac{(d-2)}{2}\int_0^1du\frac{[(1-u)(1-yu)]^{\frac{d}{2}-2}}
{(1-xu)}.
\end{equation}
Therefore, by using Eq.~(\ref{AppellF1}), one may obtain a result for the
integral $I_3^{(d)}$ in terms of the Appell function $F_1$. 
The result in terms of function $F_1$, was previously obtained in
Ref.~\cite{Fleischer:2003rm} and later on in
Ref.~\cite{Davydychev:2005nf}.
In $d=4$ space-time dimensions, the result for the integral $I_3^{(d)}$ in terms
of the function $F_3$ was given in Ref.~\cite{CabralRosetti:1998sp}.

The integral $I_3^{(d)}(0,m^2,0;~0,s_{13},s_{12})$ may also be evaluated by another
method, based on difference equations with respect to the space-time dimension
$d$.
The method of deriving dimensional recurrences is described in detail in
Refs.~\cite{Tarasov:1996br,Tarasov:2000sf}.
In the case under consideration here, we have 
\begin{eqnarray}
&&I_3^{(d+2)}(0,m^2,0;~0,s_{13},s_{12})=
\frac{2 m^2s_{13} (s_{12}-m^2-s_{13})}{(s_{12}-s_{13})^2(d-2)}
~I_3^{(d)}(0,m^2,0;~0,s_{13},s_{12})
\nonumber
\\
&& \nonumber \\
&&
~~~~~~~
+\frac{m^2}{(d-2)(s_{13}-s_{12})}~I_{11}^{(d)}(0,m^2;0)
+\frac{s_{13}(s_{12}-s_{13}-2 m^2)}{ (s_{12}-s_{13})^2(d-2)}
~I_{11}^{(d)}(0,0;s_{13}) 
\nonumber
\\
&& \nonumber \\
&&~~~~~~~
 +\frac{(m^2 s_{12}+m^2 s_{13}+s_{12} s_{13}-s_{12}^2)}
 {(s_{12}-s_{13})^2(d-2)}~I_{11}^{(d)}(0,m^2;s_{12}).
\label{dp2dp0}
\end{eqnarray}
Denoting its non-homogeneous part as $R^{(d)}$, we rewrite Eq.~(\ref{dp2dp0})
as:
\begin{equation}
I_3^{(d+2)}(0,m^2,0;~0,s_{13},s_{12})=
\frac{2m^2s_{13}(s_{12}-s_{13}-m^2)}
{(s_{12}-s_{13})^2(d-2)}~
I_3^{(d)}(0,m^2,0;0,s_{13},s_{12})
+R^{(d)}.
\end{equation}
The solution of this equation reads
\begin{equation}
I_3^{(d)}(0,m^2,0;~0,s_{13},s_{12})=
\frac{\sigma^{\frac{d}{2}}}{\Gamma\left(\frac{d-2}{2}\right)}
C_{\varepsilon}(s_{12},s_{13})-
\frac{(d-2)}{2\sigma}~\sum_{k=0}^{\infty}
\frac{\left(\frac{d}{2}\right)_k}{\sigma^{2k}}R^{(d+2k)},
\label{temp_solu}
\end{equation}
where
\begin{equation}
\sigma=\frac{m^2s_{13}(s_{12}-s_{13}-m^2)}{(s_{12}-s_{13})^2}.
\end{equation}
An arbitrary periodic function $C_{\varepsilon}(s_{12},s_{13})$ emerging in the
solution may be found from the following differential equation with respect to
the variable $s_{12}$:
\begin{eqnarray}
&&\frac{\partial}{\partial s_{12}}I_3^{(d)}(0,m^2,0;0,s_{13},s_{12})=
\frac{(d-2) (s_{13}-s_{12}+2 m^2)-2m^2}
{2(m^2+s_{13}-s_{12})(s_{13}-s_{12})}I_3^{(d)}(0,m^2,0;0,s_{13},s_{12})
\nonumber
\\
&&\nonumber \\
&&~~~~~~~~
 +\frac{(d-3)(m^2+s_{13}-2s_{12})}
{(m^2-s_{12})(m^2+s_{13}-s_{12})
 (s_{12}-s_{13}) }~I_{11}^{(d)}(0,m^2;s_{12})
\nonumber
\\
&&\nonumber \\
&&~~~~~~~~
 +\frac{(d-2)}{2(m^2-s_{12})(m^2+s_{13}-s_{12})}
 ~I_{11}^{(d)}(0,m^2;0)
\nonumber
\\
&&\nonumber \\
&&~~~~~~~~
 -\frac{ (d-3)}
{(m^2+s_{13}-s_{12})(s_{12}-s_{13})}~I_{11}^{(d)}(0,0;s_{13}).
\label{dif_equ_i3}
\end{eqnarray}
Substituting  Eq.~(\ref{temp_solu}) into Eq.~(\ref{dif_equ_i3}), we arrive at
the following equation:
%\begin{verbatim}
% diff(C(p12),p12)+(-p12+p13+3*mm)/(-p12+p13+mm)/(-p12+p13)*C(p12)
%\end{verbatim}
\begin{equation}
\frac{\partial}{\partial s_{12}} C_{\varepsilon}(s_{12},s_{13})
+\frac{(s_{12}-s_{13}-3m^2)}{(s_{12}-s_{13}-m^2)(s_{13}-s_{12})} 
C_{\varepsilon}(s_{12},s_{13})=0.
\end{equation}
Taking into account the boundary condition of the integral at $s_{12}=0$,  the
solution of this equation is
$$
C_{\varepsilon}(s_{12},s_{13})=0.
$$
Substituting explicit expressions for the propagator integrals $I_2^{(d)}$ into
Eq. (\ref{temp_solu}) leads to the following expression:
\begin{eqnarray}
&&I_3^{(d)}(0,m^2,0;~0,s_{13},s_{12})=
\nonumber
\\
\nonumber \\
&&
 - \frac{(m^2-s_{12})s_{12} + (m^2+s_{12}) s_{13} }
 {2 s_{13} (s_{13}-s_{12}+m^2)(s_{12}-m^2)}~
I_{11}^{(d)}(0,m^2;~0)~
  F_1\left( \frac{d-2}{2},\frac12,1,\frac{d}{2};
   \frac{-4 m^2 s_{12}}{(s_{12}-m^2)^2},
    -\frac{m^2}{\sigma} \right)
\nonumber \\
\nonumber \\
&&~~~~~+
 \frac{s_{13}-s_{12}}{2(s_{13}-s_{12}+m^2) s_{13}}
 I_{11}^{(d)}(0,m^2;~0)~
\Fh21\Fpo{1,\frac{d-2}{2}}{\frac{d}{2}}
\nonumber
\\
\nonumber \\
&&~~~~~
+\frac{ (s_{12} -s_{13}-2 m^2)}{2m^2(s_{13}-s_{12}+m^2)}
   I_{11}^{(d)}(0,0;~s_{13})
~\Fh21\Fmm{1,\frac{d-2}{2}}{\frac{d-1}{2}}.
\end{eqnarray}
%\begin{equation}
%z=\frac{(s_{13}-s_{12})^2}{s_{13}-s_{12}+m^2}
%\end{equation}
Comparison of the obtained result with Eq. (\ref{AppellF1}) leads to the
relation 
\begin{eqnarray}
&&F_1 \left(1,1,2-\frac{d}{2},~ \frac{d}{2};\omega,z \right)=
\frac{\omega}{2(\omega-z)(1-\omega)} 
~\Fh21\Fxy{1,\frac{d-2}{2}}{\frac{d}{2}}
\nonumber
\\
&&
+\frac{(\omega+z\omega-2z)}{2 (\omega-z)(1-\omega)(1-z)}
~F_1\left(\frac{d-2}{2},1, \frac12,\frac{d}{2}; 
\frac{\omega^2}{(\omega-z)(\omega-1)},
\frac{-4z}{(z-1)^2}\right).
\label{F1_quadratic}
\end{eqnarray}
The arguments of the function $F_1$ on the right-hand side of
Eq.~(\ref{F1_quadratic}) are connected with the arguments of the $F_1$ function
on the left-hand side by a quadratic transformation. 
Therefore, Eq.~(\ref{F1_quadratic}) is the analogue of the quadratic relation
for the Gauss hypergeometric function $_2F_1$.  
To the best of our knowledge, there is no such a relation in mathematical
literature.

%%%%%%%%%%%%%%%%%%%%%%%%%%%%%%%%%%%%%%%%%%%%%%%%%%%%%%%%%%%%%%%%%%%

\boldmath
\section{New relation between Appel function $F_1$ and
hypergeometric function $_2F_1$}
\unboldmath
\label{sec:five}

\begin{figure}[h]
\begin{center}
\includegraphics[scale=0.9]{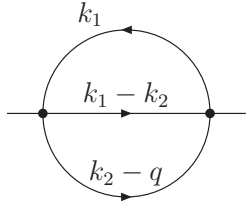}
\end{center}
\caption{Feynman diagram corresponding to the integral $J_3^{(d)}$.}
\end{figure}
%%%%%%%%%%%%%%%%%%%%%%%%%%%%%%%%%%%%%%%%%%%%%%%%%%%%%%%%%%%%%%%%%%%
In this section, we find a relation between the $F_1$ and $_2F_1$ functions by
comparing the results evaluated by two different methods for the imaginary part
of the integral
\begin{equation}
J_3^{(d)}
\equiv\int\!\!\int\frac{\rd^dk_1\rd^dk_2}{(i \pi^{d/2})^2}
\frac{1}{(k_1^2-m^2)
        ((k_1-k_2)^2-m^2)
	((k_2-q)^2-m^2) }.
\label{J3def}
\end{equation}
\vspace{1.5cm}
\noindent
The Feynman diagram corresponding to this integral is presented in Figure~3.
In Ref.~\cite{Tarasov:2006nk}, the difference equation with respect to $d$ for
the master integral $J_3^{(d)}$ was derived.
From this equation, the difference equation for the imaginary part of
$J_3^{(d)}$ may be obtained.
It reads: 
%%%%%%%%%%%%%%%%%%%%%%
\begin{eqnarray}
12 x^2 (d+1) (d-1) (3 d+4) (3 d+2)&&{\rm Im} J_3^{(d+4)}
\nonumber 
\\
-4 m^4 (x^2-3) (x^4-42 x^2+9)  (d-1) d&&{\rm Im} J_3^{(d+2)}
\nonumber 
\\
-4 m^8(x^2-1)^2 (x^2-9)^2&&{\rm Im} J_3^{(d)} = 0, 
\label{j3difference}
\end{eqnarray}
where $x=q/m$.
The solution of this equation for the imaginary part of $J_3^{(d)}$ was
presented in Ref.~\cite{Tarasov:2006nk} and reads:
\begin{equation}
{\rm Im} J_3^{(d)} = 
\frac{-4 ~\pi^2 \sqrt{3}~ m^{2d-6}}
{\Gamma\left(d-1\right)(x^2+3)}
\left[\frac{(x^2-9)^2}{27}\right]^{\frac{d-2}{2} }
 \Fh21\FXX{\frac13,\frac23}{\frac{d}{2} }.
\label{Im_part_J3}
\end{equation} 
An analytic expression for ${\rm Im} J_3^{(d)}$ may also be obtained by using
another method.
In Ref.~\cite{Bauberger:1994zz}, a one-fold integral representation for the
imaginary part of the two-loop sunrise integral with arbitrary masses was
derived.
For our case, where the masses of all propagators are the same, the imaginary
part on the cut is given in Ref.~\cite{Bauberger:1994zz} and reads:
\begin{equation}
{\rm Im} J_3^{(d)} =
\frac{-\pi}{(q^2)^{\frac{d}{2}-1}} 
\frac{\Gamma^2\bigl(\frac{d-2}{2}\bigr)}
{\Gamma^2\bigl(d-2\bigr)}
\int^{(q-m)^2}_{4m^2} \frac{d\theta }{\theta^{\frac{d}{2}-1}}
\bigl( \Lambda(\theta,m^2,m^2) \Lambda(\theta,q^2,m^2)\bigr)^{\frac{d-3}{2}},
\label{Im_p_Bauberger}
\end{equation}
where $\Lambda(x,y,z)$ is defined in Eq.~(\ref{eq:lambda}).
Changing the integration variable in Eq.~(\ref{Im_p_Bauberger}) as
$\theta= 4m^2+(q-3m)(q+m)\beta$
%\begin{equation}
%\theta= 4m^2+(q-3m)(q+m)\beta
%\end{equation}
leads to the following expresssion:
\begin{eqnarray}
&&{\rm Im} J_3^{(d)} =
\frac{-\pi}{(q^2)^{\frac{d}{2}-1}} 
\frac{\Gamma^2\bigl(\frac{d-2}{2}\bigr)}
{\Gamma^2\bigl(d-2\bigr)}
\frac{(q-3m)(q+m)}{2m} 
\bigl[(q-m)(q+3m)(q+m)^2(q-3m)^2\bigr]^{\frac{d-3}{2}}
\nonumber \\
&&~~\times
\int^{1}_{0} d\beta 
\Bigl\{ \beta(1-\beta)
\bigl[1-\frac{(q+m)(q-3m)}{(q-m)(q+3m)} \beta \bigr]
 \Bigr\}^{\frac{d-3}{2}}
\frac{1}{ \bigr[1+\frac{(q+m)(q-3m)}{4m^2}\beta \bigr]^{\frac{1}{2}}}.
\label{Im_p_f1}
\end{eqnarray}
As follows from Eq.~(\ref{F1_int_repr}), the integral on the right-hand side of
Eq.~(\ref{Im_p_f1}) is proportional to the Appell function $F_1$.
Using Eq.~(\ref{F1_int_repr}) and comparing Eqs.~(\ref{Im_p_f1}) and
(\ref{Im_part_J3}), we arrive at the following relation:
\begin{eqnarray}
&&F_1\left( \frac{d-1}{2},\frac{3-d}{2},\frac12,d-1;
\frac{(x+1)(x-3)}{(x-1)(x+3)},
-\frac{(x+1)(x-3)}{4} \right)
\nonumber \\
&&
=\frac{2\sqrt{3}}{(x^2+3)}\left[\frac{16}{27}\frac{x^2(x+3)^2}{(x+1)^2}
\right]^{\frac{d-2}{2}}
\left[(x+3)(x-1)\right]^{\frac{3-d}{2}}
\Fh21\FXX{\frac13,\frac23}{\frac{d}{2}}.
\label{F1viaF21}
\end{eqnarray}

It is interesting to note that, at $d=2$, the Appell function $F_1$ in
Eq.~(\ref{F1viaF21}) may be expressed in terms of the ${}_2F_1$ function
with the help of the equation
(see, for example, Ref.~\cite{Prudnikov:1986})
\begin{equation}
F_1(a,b,b',b+b',w,z)=(1-z)^{-a}\Fh21\Fwz{a,b}{b+b'}.
\end{equation}
This leads to the relation:
\begin{equation}
\Fh21\Fbb{\frac12,\frac12}{1}=\frac{\sqrt{3(x+3)(x-1)^3 }}
{(x^2+3)}
\Fh21\FXX{\frac13,\frac23}{1}.
\label{Ramanujan_type_relation}
\end{equation}
The hypergeometric function $_2F_1$ on the left-hand side of this equation
is proportional to the complete elliptic integral of the first kind.
Relations between hypergeometric functions with parameters $1/2,1/2,1$ and
$1/3,2/3,1$ but with arguments different from that in
Eq.~(\ref{Ramanujan_type_relation}) were first derived by Ramanujan 
in Ref. \cite{Ramanujan:1913}. 

%%%%%%%%%%%%%%%%%%%%%%%%%%%%%%%%%%%%%%%%%%%%%%%%%%%%%%%%%
%%%%%%%%%%%%%%%%%%%%%%%%%%%%%%%%%%%%%%%%%%%%%%%%%%%%%%%%%
%%%%%%%%%%%%%%%%%%%%%%%%%%%%%%%%%%%%%%%%%%%%%%%%%%%%%%%%%
\section{Conclusions}
\label{sec:six}

In this section, we briefly summarize the most important results obtained in
this paper and point out some topics which may be of interest for future
investigations.
Specifically,
\begin{itemize}
\item  a new analytic expression for the one-loop propagator-type Feynnman
integral was derived;
\item a new formula transforming the Appell function $F_1$ to the
hypergeometric function $_3F_2$ was presented;
\item a new formula transforming the Appell function $F_4$ to a combination of
two $F_1$ functions was found;
\item a new formula connecting the Kamp\'e de F\'eriet function and the Appell
function  $F_1$  was obtained;  
\item a formula for the quadratic transformation of the Appell function $F_1$
was given;
\item a new relation between the Appell function $F_1$ and the elliptic-type
Gaussian hypergeometric function $_2F_1$ was established.
%\item
%for the Appell function $F_1$ quadratic transformation
%is found
\end{itemize}

We expect that the formulae found for the abovementioned hypergeometric
functions will be useful, in particular, for finding relations between 
transcendental numbers.
Such relations may be found, for example, via the expansions in
$\varepsilon=(4-d)/2$ of the presented results.
These relations may also be useful for simplifying the $\varepsilon$ expansions
of hypergeometric functions involved because some of them may have simpler
integral representations than the others. 

\section*{Acknowledgments}

This research was supported in part by the German Federal Ministry for Education
and Research BMBF through Grant No.\ 05~HT6GUA, by the German Research
Foundation DFG through the Collaborative Research Centre No.~676
{\it Particles, Strings and the Early Universe---The Structure of Matter and
Space Time}, by the Helmholtz Association HGF through the Helmholtz
Alliance Ha~101 {\it Physics at the Terascale}, and by the National Science
Foundation NSF under Grant No.\ NSF~PHY05--51164.

%%%%%%%%%%%%%%%%%%%%%%%%%%%%%%%%%%%%%%%%%%%%%%%%%%%%%%%%%

\end{document}